\documentclass[reprint,secnumarabic,amssymb,nobibnotes, aps,prb]{revtex4-1}
\usepackage{graphicx}
\usepackage{amsmath}
\usepackage{subfigure}
\usepackage{verbatim}
\usepackage[english]{babel}
\usepackage[utf8]{inputenc}
\usepackage{dsfont}
\usepackage[colorinlistoftodos]{todonotes}
\usepackage[version=3]{mhchem}
\usepackage{chemfig}
\usepackage{cancel}
\usepackage{hyperref}
\usepackage{tabularx}
\usepackage{xcolor}
\usepackage{harpoon}
\usepackage{upgreek}
\usepackage{epstopdf}
\usepackage{bm}
\usepackage{color}

\graphicspath{{./figs/}}

\makeatletter
\newsavebox{\@brx}
\newcommand{\llangle}[1][]{\savebox{\@brx}{\(\m@th{#1\langle}\)}%
  \mathopen{\copy\@brx\kern-0.5\wd\@brx\usebox{\@brx}}}
\newcommand{\rrangle}[1][]{\savebox{\@brx}{\(\m@th{#1\rangle}\)}%
  \mathclose{\copy\@brx\kern-0.5\wd\@brx\usebox{\@brx}}}
\makeatother

\begin{document}
\title{
Theoretical study for 3D quantum Hall effect in a periodic electron system
}

\author{H. Geng$^{1,2}$}
\author{G. Y. Qi $^{1}$}

\author{L. Sheng$^{1,2}$}
\email{shengli@nju.edu.cn}

\author{W. Chen$^{1,2}$}
\email{pchenweis@gmail.com}
\author{D. Y. Xing$^{1,2}$}

\affiliation{$^{1}$ National Laboratory of Solid State Microstructures and Department of
	Physics, Nanjing University, Nanjing 210093, China\\
	$^{2}$ Collaborative Innovation Center of Advanced Microstructures, Nanjing
	University, Nanjing 210093, China}
\date{\today }

\begin{abstract}
The exsitance of three-dimensional Hall effect (3DQHE) due to spontaneous
Fermi surface instabilities in strong magnetic field was proposed decades ago,
and has stimulated recent progress in experiments.
The reports in recent experiments show that 
the Hall plateaus and vanishing transverse magneto-resistivities (TMRs) 
(which are two main signatures of 3DQHE)
are not easy to be observed in natural materials.
And two main different explanations of the slow varying slope like Hall plateaus and non-vanishing 
 TMRs (which can be called as quasi-quantized Hall effect (QQHE)) have been proposed.
By studying the magneto-transport with a simple effective periodic 3D system, 
we show how 3DQHE can be achieved in certain parameter regimes at first.
We find two new mechanisms that may give rise to QQHE. 
One mechanism is the "low" Fermi energy effect, and the other is the "strong" impurity effect.
Our studies also proved that the artificial superlattice is an ideal platform for realizing 3DQHE
with high layer barrier periodic potential.
\end{abstract}

\maketitle
\section{INTRODUCTION}
The discoveries of quantum Hall effects (QHE) \cite{QHEexp1, QHEexp2,QHEGraphene}
in the two-dimensional (2D) electron gas have inspired the discovery and
classification of topological materials
in condensed-matter physics\cite{TI1RevModPhys.82.3045,TI2RevModPhys.83.1057, TS1RevModPhys.90.015001}.
The main feature of 2DQHE is that the quantized Hall conductivities take values
$\sigma_{xy} = \nu e^2/h$ with $\nu$ integers
and the transverse TMCs  $\sigma_{\alpha \alpha}$ or
TMRs  $\rho_{\alpha \alpha}$ with $\alpha = x,y$ vanish when the Hall plateaus appear.
Here the word "transverse" means perpendicular to the direction of the magnetic field, e.g. $z$ direction.
Another important feature of 2DQHE is
the formation of dissipationless one-dimensional (1D) chiral edge states protected by topology\cite{QHETKNN}.
Similar effects have been discovered in
quasi-2D systems with stacking 2D QHE layers
where the interlayer coupling is much weaker than the Landau level (LL) spacing
\cite{Qusi2DQHEExp1, Qusi2DQHEExp2,Qusi2DQHEExp3,Qusi2DQHEExp4, Qusi2DQHEExp5masuda2016quantum}.
These quasi-2DQHE have similar quantized Hall conductance $G_{xy} = \nu e^2 /h$
which can be regarded as 2DQHE in quasi-2D systems.
In contrast to 2D and quasi-2D electron gas systems,
the QHE was thought to be forbidden  in three dimensional(3D) electron gas.
The reason is that the third dimension,
along the direction of the magnetic field,
spreads the LLs into overlapping bands,
thereby no energy gap exists between the LLs,
and the band quantization is destroyed.
But the discoveries of 3D topological systems give a possible way to realize QHE.
The key point is that some 3D topological systems may have topologically protected
2D electron gas at the surfaces\cite{TI1RevModPhys.82.3045, TI1RevModPhys.82.3045, TS1RevModPhys.90.015001}
which can give rise to 2DQHE\cite{QHETI, QHEWSPhysRevLett.119.136806, QHEWSLHZNSR, QHENLSLHZ}.
This may also result in quantized Hall conductance just as quasi-2D systems.

Despite searching for quantized Hall conductances in quasi-2D systems or topological 3D systems,
    people also try to search for the quantized Hall conductivities in bulk 3D systems.
    In fact, in the vicinity of the quantum limit, 3D electron systems are also tend to form varieties of correlated
    electron states\cite{3dqheThe1Haperin}, including Luttinger liquids, charge density waves (CDWs), 
    spin density waves (SDWs), valley density waves(VDWs),  excitonic insulators, Wigner crystals, 
    Hall crystals, and staging transitions in the case of highly anisotropy layered systems
    \cite{Celli1965,3dqheThe1Haperin, MacDonald1988, Takada1998, Chalker1999,Burnell2009,Akiba2015}.
    It has been predicted that 3DQHE could  be observed in semimetals 
    and doped semiconductors\cite{3dqheThe1Haperin, Kohmoto1992, BernevigPRL2007,Koshino2003prb}.
    In these systems, the application of  a magnetic field would lead to Fermi surface instability,
    which may cause a periodic modulation of the electron density  like CDWs, or SDW along the direction of the magnetic field.
    From this point, signatures of 3DQHE are also manifestations of the emergence of correlated states.
The main distinct signature of 3DQHE
is the value of Hall conductivities plateaus
$\sigma_{xy} = G_z e^2/2 \pi h$ which are different
from 2DQHE, here $G_z$ is the z-component of a reciprocal (super)lattice vector
or the period of $z$ directly potential.
The 3DQHE is also expected the same signature of vanishing TMCs
, namely $\sigma_{xx}=0$.
The dissipationless edge states in 3DQHE systems
perform differently compared with edge states in 2DQHE systems.
The surface parallel to the magnetic field may carry surface states that is
dissipationless along the transverse direction but diffusive
along the longitudinal $z$ direction\cite{3DQHESS1PhysRevLett.76.2782,3DQHESS2PhysRevLett.75.4496}.

    Inspired by these ideas, signatures of 3DQHE have been reported to be observed in several systems.
    Decades ago, the bulk QHE was observed in fabricated artificial superlattices\cite{3DQHEPeriodic1986}.
    Even though it is not a strict 3DQHE as the Hall resistivities is still dependent on the thickness of the systems
    rather than only depend on the thickness of only one layer of the superlattice,
    it's still a good example that demonstrates the exsitance of periodic potential will open gaps
    in the overlapping LLs.
    We can expect that when the thickness of the superlattice is large enough, the true 3DQHE may show up.
    On the other hand,the spontaneous mechanisms to get
    3DQHE systems were proposed theoretically decades ago,
    but this novel phenomenon has not been observed experimentally
    until recently\cite{3dqheExp1, 3dqheExp2,Galeski2020} in the semimetal systems of   
    bulk $\text{ZrTe}_5$ and $\text{HfTe}_5$ .
    The 3DQHE observed in experiment\cite{3dqheExp1} is carefully explained by
    the CDW mechanism\cite{LHZCDWQHE}.
    Particularly, the Hall resistivity $\rho_{xy}$ has been found to 
    exhibit a plateau with a value of $h\pi/e^2  k_F$ which is in consistent with the CDW mechanism.
    This is because the Fermi surface instability lead to the the formation of the CDW with a wavelength of 
    half of the Fermi wave length, and specifically, the period of the CDW  is  $Z_{\text{CDW}}=\pi/k_F$.
    However, the other signature of 3DQHE, 
    vanishing TMCs are only reported to be observed in the paper\cite{3dqheExp1}.
    And in other papers, TMCs (TMRs) are minimum finite values, and they think may be other Fermi pockets near
    the Fermi surface lead to this\cite{{3dqheExp2}}.
    Very recently, another group\cite{3DQQHEZrTe5} also investigate the physics of $\text{ZrTe}_5$ in magnetic field.
    They claimed that they observed the so-called quasi-quantized Hall effect (QQHE) in $\text{ZrTe}_5$ systems.
    The main difference between QQHE and 3DQHE is that in QQHE, the quasi-quantized
    Hall plateaus show up with nonvanishing TMCs (or TMRs) rather
    than the vanishing ones.
    Above all, it seems that the tranpsort signatures of 3DQHE in electron systems with periodic potential
    are rather complex.
    It is necessary to theoretically study the TMCs (or TMRs) and Hall conductivities (or resistivities)
    in detail and give more insight into this novel phenomenon.

    In this paper,
    we effectively model these periodic systems with
    periodic square-well potential along $z$ direction
    which is usually known as the Kronig-Penney model.
    Then, a magnetic field along $z$ direction is applied in the system.
    We explicitly calculate
    the energy bands, DOS, Hall conductivities without impurities.
    From these results, the origin of 3DQHE can be clearly understood
    by the gaps origin from the periodic potential.
    To compare with experimental results, impurities
    are taken into consideration, and we studied
    the TMRs $\rho_{xx}$ and Hall resistivities $\rho_{xy}$.
    Our results show that quantized Hall resistivities $\rho_{xy}$
    and the vanishing TMRs $\rho_{xx}$ are surely a signature of
    the appearence ot 3DQHE in periodic system.
    And, what's more, we find that even results with
    slope-like quasi-quantized Hall resistivities with finite 
    TMRs (which we can call QQHE in this paper) 
    in the vicinity of the quantum limit can have other origins 
    despite the  reansons metioned in papers\cite{3dqheExp2,3DQQHEZrTe5}.
    We find the "low" Fermi energies and "strong" impurities may give rise to 
    QQHE in periodic potential electron systems.
    We cannot just conclude whether periodic potential exists
    from QQHE observations.
    Finally, the advantage of the artificial superlattice is also discussed.

This article will be arranged in the following structure.
In section II, the general model and Methods are presented.
In section III, the Kronig-Penney model for periodic potential
along $z$ direction has been considered,
and the main results will be presented.
In section IV, 
    a discussion of the results and  a conclusion will be made.

\section{Model And Methods}
To model the periodic system made by superlattice or modulated periodic potential,
let's consider a cuboid 3D normal metal system with length,
width and height being $L_x$, $L_y$, $L_z$ respectively.
The hamiltonian, including vector potential $\mathbf{A}$,
modulation periodic potential $V(z)$ satisfying $V(z+Z)=V(z)$
and random potential $U(\mathbf{r})$, can be described as
\begin{equation}\label{ham1}
    H = H_0+U(\bm{r})~,
\end{equation}
where $H_0=\frac{\bm{P}^2}{2M}+V(z)$ is the hamiltonian of free electron
in the magnetic field with the mechanical momenta being
 $\bm{P}=\bm{p}+e\bm{A}$ ,
$M$ is the effective mass assumed isotropic
and the canonical momenta $\bm{p}=-i \hbar {\bm{\nabla}}$.
Although what we are considering is a rough and simple approximation,
it is enough to give some insight into the real systems.

\subsection{Landau Levels and Quantized Hall conductivities}
Applying the magnetic field along the $z$ direction, namely, $\mathbf{B}=\left(0, 0, B\right)$
and choosing the vector potential within the Landau gauge as $ \mathbf{A} = \left(0, Bx, 0\right)$,
we can write down the basic commutation relations for the coordinate and momentum operators,
$\left[ P_x, P_y \right] = -i e\hbar B$ and
$\left[ \mathbf{P}, f\left( \bm{r} \right) \right] = -i \hbar \bm{\nabla} f\left( \text{r} \right)$,
here $f\left( \bm{r} \right)$ is a function of coordinate $\bm{r}$ .
The velocity operator can be obtained as
$\bm{v} = \nabla_{\bm{P}}H =\frac{\bm{P}}{M}$.
The stationary Schr\"{o}dinger equation of $H_0$ is
\begin{equation}\label{Seq}
    H_0\left |\Psi\right> = E\left |\Psi\right>.
\end{equation}
One way to solve this problem is to
introduce the ladder operators $a$ and $a^+$
with commutation relation $\left[a,a^+\right] = 1$.
Then the components of mechanical momentum can be expressed as
$P_x = \frac{\hbar}{\sqrt{2}l_B}\left(a+a^+\right)$ and
$P_y = \frac{i\hbar}{\sqrt{2}l_B}\left(a-a^+\right)$,
where $l_B = \sqrt{\frac{\hbar}{eB}}$ is the magnetic length.
And similarly, the velocity operators are
\begin{equation}\label{vel2}
    \begin{split}
        \hat{v}_x = \frac{\hbar}{\sqrt{2}M l_B}\left(a+a^+\right),\\
        \hat{v}_y = \frac{i\hbar}{\sqrt{2}M l_B}\left(a-a^+\right).
    \end{split}
\end{equation}

The hamiltonian $H_0$ is expressed in terms of ladder operators as
\begin{equation}\label{H0Ladder}
    H_0 = \hbar \omega_c \left( a^+a +\frac{1}{2} \right) + \frac{p_z^2}{2M}+V\left(z\right),
\end{equation}
and here $\omega_c=\frac{eB}{M}$ is the cyclotron frequency.
As can be easily seen from Eq.(\ref{H0Ladder}), $H_0$ can be
divided into two independent parts $h^{\parallel}$ and $h^{\bot}$, which are
\begin{equation}\label{hpar}
    h^{\parallel} = \frac{p_z^2}{2M}+V\left(z\right),
\end{equation}
and
\begin{equation}\label{hbot}
    h^{\bot} = \hbar \omega_c \left( a^+a +\frac{1}{2} \right).
\end{equation}
Therefor, we can get the eigenstates of $h^{\parallel}$ and $h^{\bot}$ respectively,
and the eigenstates of $H_0$ is just can be given by them.
From standard quantum mechanics textbook, e.g.\cite{landau1998course},
the eigenstate of $h^\bot$ can be represented by two quantum numbers:
the LL $N$ and the guiding center $X$,
which are the eigenvalues of $a^+ a$ and $-\frac{l_B^2}{\hbar} p_y$.
And the eigenstate of $h^{\parallel}$ is just one dimensional Bloch waves,
which can also be denoted by two quantum numbers,
the Bloch band number $n_b$ and the quasi wave number $k_z$.
The allowed values that can be taken by these quantum numbers are
$N\in \left\{0, 1, 2, \cdots \right\}$,
$X \in \left\{x|0<= x<L_x\right\}$,
$n_b \in \left\{1,2,3,\cdots\right\}$,
and $k_z \in \left\{k_z| -\frac{\pi}{Z}<k_z<= \frac{\pi}{Z}\right\}$.

Denoting the complete set of quantum numbers
as $\gamma=(N,X, n_b, k_z)$ for convenience,
we can acquire the eigensolution of $H_0$:
the eigenstate can be formally marked as $\left|\gamma \right>$
and the corresponding eigenenergy is
\begin{equation}\label{eigenergy}
    E_{\gamma} = \left(N +\frac{1}{2} \right) \hbar \omega_c + E_{n_b, k_z}.
\end{equation}
In this eigen basis, the velocity operators can be represented as

\begin{equation}\label{velLL}
    \begin{split}
        \hat{v}^x_{\gamma,\gamma'} = \frac{\hbar \delta_{ \overline{\gamma},  \overline{\gamma}'}}
        {\sqrt{2}M l_B}\left(\sqrt{N+1}\delta_{N+1, N'}
        +\sqrt{N'+1}\delta_{N, N'+1}\right),\\
        \hat{v}^y_{\gamma, \gamma'} = \frac{i\hbar\delta_{ \overline{\gamma},  \overline{\gamma}'}}
        {\sqrt{2}M l_B}\left(\sqrt{N+1}\delta_{N+1, N'}
        -\sqrt{N'+1}\delta_{N, N'+1}\right),
    \end{split}
\end{equation}
here, $\overline{\gamma}=(X, n_b, k_z)$, and we use the relations
$a^+\left|N, \overline{\gamma}\right>=\sqrt{N+1}\left|N+1,\overline{\gamma}\right>$,
 $a\left|N,\overline{\gamma}\right>=\sqrt{N}\left|N-1, \overline{\gamma}\right>$,
and $\left<N, \overline{\gamma}|N', \overline{\gamma}'\right>=\delta_{N, N'}\delta_{ \overline{\gamma},  \overline{\gamma}'}$.

% \subsection{Kubo formula for Hall conductivity}
To obtain the Hall conductivity,
we mainly use the Kubo-Greenwood formula
which can be formally experessed as \cite{ZieglerKG2GreenPRL, Kristina2017}
\begin{equation}\label{Kubo1}
    \sigma_{ij} =
   -\frac{i  g_{s} e^2 \hbar}{\Omega}
    \sum_{\varepsilon_{\alpha}\neq \varepsilon_{\beta}}
    \frac{f\left(\varepsilon_{\alpha}\right)-f\left(\varepsilon_{\beta}\right)}{\varepsilon_{\alpha}-\varepsilon_{\beta}}
    \frac{
    \left< \alpha \right| \hat{v}_i\left|\beta\right>
    \left< \beta \right| \hat{v}_j\left|\alpha\right>
    }{
    \varepsilon_{\alpha}-\varepsilon_{\beta}-\hbar\omega+i \eta
    },
\end{equation}
here, $\varepsilon_{\alpha}$ and $\varepsilon_{\beta}$ are the eigenenergies
corresponding to the eigenstates $\left|\alpha \right>$ and $\left|\beta \right>$ of the system respectively,
and $g_s$ is a degeneracy factor.
$\Omega=L_x L_y L_z$ is the volume of the system.
$f\left(\varepsilon_{\alpha}\right)$ and $f\left(\varepsilon_{\beta}\right)$ are the Fermi-Dirac distribution functions,
defined as $f(x) = 1/\left[e^{(x-E_F )/k_B T} + 1\right]$
where $E_F$ is the Fermi Energy of the system.
$\hat{v}_i$ and $\hat{v}_j$ are the velocity operators
and $\eta$ is the small positive value which can be regarded as the selfenergy
 arising from defects.

The Hall conductivities at zero temperature, namely, $\omega \rightarrow{0}$,
and for clean system, $\eta \rightarrow{0^+}$, Eq.(\ref{Kubo1}) can be further simplified as
\begin{equation}\label{Kubo2}
    \sigma_{x y} =
    \frac{2 g_s  e^2 \hbar}{\Omega}
     \sum_{N, N',\overline{\gamma}}
    \frac{ \text{Im} \left[\left<N',\overline{\gamma}\right| \hat{v}_x\left|N, \overline{\gamma}\right>
    \left< N,\overline{\gamma} \right| \hat{v}_y\left|N', \overline{\gamma}\right>\right]
    }{
    \left( E_{N',\overline{\gamma}}-E_{N,\overline{\gamma}}\right)^2
    },
\end{equation}
here, $E_{N',\overline{\gamma}}<E_F<E_{N,\overline{\gamma}}$,
and $\text{Im}\left[c \right]$ is taking the imaginary part of some complex number $c$.
Firstly, substitute the matrix elements of velocities in Eq. (\ref{velLL}) into the Eq. (\ref{Kubo2}).
Secondly, for the same $\overline{\gamma}$,
    the eigenenergies only differ by the energy of Landau Levels.
    That is, $E_{N',\overline{\gamma}}-E_{N,\overline{\gamma}}=
    E^{\bot}_{N',X}-E^{\bot}_{N, X}=(N'-N)\hbar \omega_c$.
Thirdly, the degeneracy of LLs is $\sum_{X} 1 = \frac{L_x L_y}{2\pi l_B^2}$.
Finally, with periodic boundary condition in $z$ direction,
$\sum_{k_z} \cdots \rightarrow{\frac{Lz}{2\pi}\int}dk_z\cdots$ .
And finally, we get the expression for the Hall conductivity
\begin{equation}\label{HallCond0}
    \sigma_{x y} = g_s\frac{e^2}{2 \pi h} \left(\sum_{N,n_b} \int_{E<E_F} 1 dk_z \right).
\end{equation}
By taking $g_s=1$ for non-degeneracy case, the result is in consistent with Halperin's result\cite{3dqheThe1Haperin}.

% \section{Quantized Hall conductivity}
To get stable 3D QHE from the system, the gap of Landau Levels should be large enough,
so that the first Landau Level Bloch band should be larger
than the zeroth Landau Level first Bloch band.
Or say, $E(N=0, X,n_b=1, k_z=\pm\pi/Z)<E(N=1, X, n_b=1, k_z=0)$,
which also means that the gap of the nearest Landau Levels is larger
than the energy difference in the first Bloch bands.
In this case, there will be at least one stable quantized Hall plateau.

\subsection{Impurities, level broadening, and magneto-conductivities}

To gain a further understanding 3DQHE system,
we need to get the vanishing TMCs accompanied by the Hall plateaus.
For the transverse electronic transport, namely, $\mathbf{E}\bot \mathbf{B}$,
noticing that the LLs doesn't disperse along $x$ or $y$ direction from Eq.\eqref{eigenergy},
electrons described by the hamiltonian $H_0$ can't drift in the $x-y$ plane.
More specifically,
we look back to Eq.\eqref{Kubo1} and simplify it
by taking subscripts $j=i$ with $\hbar \omega\rightarrow 0$, so
\begin{equation}\label{LongCond}
    \sigma_{ii} =
   -\frac{\pi  g_{s} e^2 \hbar}{\Omega}
    \sum_{\varepsilon_{\alpha}\neq \varepsilon_{\beta}}
    \frac{\partial f\left(\varepsilon_{\alpha}\right)}{\partial\varepsilon_{\alpha}}
    \delta\left( \varepsilon_{\alpha}-\varepsilon_{\beta} \right)
     \left| \left< \alpha \right| \hat{v}_i\left|\alpha\right> \right|^2~.
\end{equation}
Combining Eq.\eqref{LongCond} with Eq.\eqref{velLL},
we can see that the diagonal elements of the velocity terms of $i=x$ and $i=y$ will vanish without impurities.

It is time that we pull in
the random potential $U\left( \mathbf{r} \right)$ induced by impurities and defects.
% \sout{to get the nonvishing TMCs.}
Let's consider randomly distributed potential
\begin{equation}\label{RandomPotential}
    U\left( \mathbf{r} \right) =  \sum_{i=1}^{N_{\text{imp}}} U_i u\left( \mathbf{r} -\mathbf{R}_i  \right),
\end{equation}
where $U_{i}$  follows Binomial distribution with equal probability to
choose a value of $-W$ or $W$,
$\mathbf{R}_i $ is  uniform  randomly distributed in the volume $\Omega$,
and $N_{\text{imp}}$ is the number of the impurities.

For the static case $\omega\rightarrow 0$, a convenient derivation of
Kubo-Greenwood formula is the so called Kubo-Bastin formalism\cite{Bastin1971}.
Taking into account that $\lim_{\eta\rightarrow 0^+}
\frac{1}{\left( \varepsilon_{\alpha}-\varepsilon \right) \left( \varepsilon_{\alpha}-\varepsilon +i \eta\right)}=
\lim_{\eta\rightarrow 0^+} \frac{d}{d\varepsilon} \left\{ \frac{1}{\varepsilon_{\alpha}-\varepsilon +i \eta} \right\}$,
Eq.(\ref{Kubo1}) can be written as\cite{Kristina2017}
\begin{equation}\label{Kubo-Bastin}
    \begin{split}
        \sigma_{ij} &=-\frac{g_s e^2\hbar}{2\pi\Omega}\int d\varepsilon f_{F}\left(\varepsilon\right)\times\\
        &\text{Tr}\left[ \hat{v}_i\frac{\partial G^{+}_{ \varepsilon}}{\partial\varepsilon}
        \hat{v}_j\Delta G_{\varepsilon} -\hat{v}_i\Delta G_{\varepsilon}
        \hat{v}_j\frac{\partial G^{-}_{ \varepsilon}}{\partial\varepsilon}
        \right],
    \end{split}
\end{equation}
where we have used the definitions $\text{Tr}[\hat{O}]$
taking the trace of the operator matrix $\hat{O}$,
and the one-body green's function
$G\left( z \right) = \frac{1}{z-H}$,
and also the relations\cite{Economou2006}
$\delta\left( \lambda - H\right) = -\frac{1}{2 \pi i} \Delta G_{\lambda} = \mp\frac{1}{\pi} \text{Im} G^{\pm}_{\lambda}$,
 $\Delta G_{\lambda}= \left( G^+_{\lambda} - G^-_{\lambda} \right) $,
and $G^{\pm}_{\lambda} = G\left( \lambda \pm i 0^+ \right)$ with $\lambda$ being a real number.
Macroscopic physical quantities, e.g. transport conductivities,
are associated  with averaged characteristics over the distribution of impurities.
For this reason, the conductivities introduced by Eq. (\ref{Kubo-Bastin}) must be averaged,
\begin{equation}\label{KBAvg}
    \begin{split}
        \sigma_{ij} &=-\frac{g_s e^2\hbar}{2\pi\Omega}\int d\varepsilon f_{F}\left(\varepsilon\right)\times\\
        &\llangle\text{Tr}\left[ \hat{v}_i\frac{\partial G^{+}_{ \varepsilon}}{\partial\varepsilon}
        \hat{v}_j\Delta G_{\varepsilon} -\hat{v}_i\Delta G_{\varepsilon}
        \hat{v}_j\frac{\partial G^{-}_{ \varepsilon}}{\partial\varepsilon}
        \right]\rrangle,
    \end{split}
\end{equation}
where  $\llangle \cdots \rrangle$
means taking the average over the random potential.

Assuming $j=i$ , a more simple form of Eq. (\ref{KBAvg}) is obtained
for the diagonal conductivity tensor\cite{Vasko2005SDH},
\begin{equation}\label{siiAvg}
    \begin{split}
    \sigma_{ii} =&
    -g_{s}\frac{ e^2 \hbar}{4 \pi\Omega}
    \int d\varepsilon
    \left(-  \frac{ \partial f\left(\varepsilon\right) }
    {\partial \varepsilon}\right)\times\\
    &\llangle\text{Tr}\left[ \hat{v}_i \Delta G_{\varepsilon} \hat{v}_i \Delta G_{\varepsilon}  \right]\rrangle.
    \end{split}
\end{equation}

The averaged Green's function $G^{s}_{\varepsilon}$ (with $s=\pm$) of the
electron interacting with a random scattering potential has only none-zero diagonal
elements in the LLs representation. This can be expressed as
\begin{equation}\label{}
    \llangle G^{s}_{\varepsilon}\left( \gamma, \gamma' \right) \rrangle =
G^{s}_{\varepsilon}\left( \gamma\right)\delta_{\gamma, \gamma'}.
\end{equation}
Below we consider the limit of short-range scattering potential,
and neglect the vertex correction to the correlation functions in Eq. \eqref{KBAvg}.
Combining Eqs. \eqref{Kubo-Bastin}, \eqref{KBAvg} and \eqref{velLL}, we obtain TMC
\begin{equation}\label{sxx_lb}
    \begin{split}
        \sigma_{xx} = &g_s \frac{e^2 \left( \hbar\omega_c \right)^2}{4\pi^2 h}
        \int d \varepsilon \left(- \frac{ \partial f }
    {\partial \varepsilon}\right) \int d k_z \sum_{N, n_b} \left( N+1 \right)\times\\
        & \text{Re}\left[ G^-_{\varepsilon}\left( n_b, k_z, N \right)
        \Delta G_{\varepsilon}\left( n_b, k_z, N+1 \right)
        \right],
    \end{split}
\end{equation}
and the Hall conductivity
\begin{equation}\label{sxy_lb}
    \begin{split}
        \sigma_{xy} = &g_s \frac{e^2 \left( \hbar\omega_c \right)^2}{4\pi^2 h}
        \int d \varepsilon \int d k_z \sum_{N=0}^{\infty}\sum_{n_b} \left( N+1 \right)\times\\
        & \text{Im}\left[ \frac{\partial f}{\partial \varepsilon}
        G^-_{\varepsilon}\left( n_b, k_z, N \right)
        G^{+}_{\varepsilon}\left( n_b, k_z, N+1 \right)+ \right.\\
        &f\left(\varepsilon\right)\left(
         G^+_{\varepsilon}\left( n_b, k_z, N \right)
        \frac{\partial G^{+}_{ \varepsilon}\left( n_b, k_z, N+1 \right)}{\partial\varepsilon}-\right.\\
        &\left. \left.
        \frac{\partial G^{+}_{ \varepsilon}\left( n_b, k_z, N \right)}{\partial\varepsilon}
        G^{+}_{ \varepsilon}\left( n_b, k_z, N \right)
        \right) \right]
    \end{split}
\end{equation}

In the self-consistent Born approximation(SCBA)\cite{Vasko2005SDH,ando1974theory1, ando1974theory2},
we get
\begin{equation}\label{scba_selfenergy}
    \Sigma_{\varepsilon}^s\left( \gamma \right)
    =\sum_{\gamma'} \llangle  \left| U_{\gamma',\gamma}\right|^2 \rrangle
    G^{s}_{\varepsilon}\left(\gamma'\right),
\end{equation}
with
\begin{equation}\label{green_SCBA}
    G^{s}_{\varepsilon}\left(\gamma\right)
    =\left[\varepsilon - E_{\gamma}- \Sigma^{s}_{\varepsilon}\left(\gamma\right)\right]^{-1}.
\end{equation}
Here, %\sout{$\gamma$ stands for eig-state of $H_0$ indexed by $(N, X, n_b, k_z)$ ,}
$U_{\gamma',\gamma} = \left< \gamma' \right| U \left| \gamma\right>$
is the matrix elements of the potential $U$ in the $H_0$ eigenbasis.
Let's transform the impurity potential into  Fourier series according to standard relations
\begin{equation}\label{impurityFFTEqs}
    \begin{split}
        u\left( \mathbf{r} \right) = \frac{1}{\Omega}\sum_{\mathbf{q}} u_{\mathbf{q}} e^{i\mathbf{q}\cdot \mathbf{r}}\\
        u_{\mathbf{q}}=  \int d \mathbf{r}  u\left( \mathbf{r} \right) e^{-i\mathbf{q}\cdot \mathbf{r}}
    \end{split}
\end{equation}
The impurity averaged binary potential correlation function is\cite{Vasko2005SDH}
\begin{equation}\label{impCorrelationEq}
    \llangle U\left( \mathbf{r} \right) U\left( \mathbf{r}^{\prime} \right)  \rrangle=
    \frac{1}{\Omega} \sum_{\mathbf{q}} \left| u_{q}\right|^2 n_{\text{imp}}W^2
    e^{i \mathbf{q}\cdot \left( \mathbf{r}-\mathbf{r}^{\prime} \right)},
\end{equation}
where $ n_{\text{imp}}$ is the impurity concentration, and
we can see the Fourier component of the averaged correction function is
$w\left( \mathbf{q} \right)= \left| u_{q}\right|^2 n_{\text{imp}}W^2$.
To obtain Eq. \eqref{scba_selfenergy}, we need to use the following matrix element
\begin{equation}\label{UmatEq}
    \begin{split}
        &\left| \left< \gamma \left|e^{i\mathbf{q}\cdot\mathbf{r}}\right|\gamma^{\prime} \right> \right|^2=
        \Phi_{N_2 N_1}\left( \frac{q_{\bot}l_B^2}{2} \right)\delta_{k_y, k_y^\prime+q_y}\times\\
        &~~~~~~~~~~~~~~\left| \left< n_b, k_z \left|e^{iq_{z}z}\right|n_b^{\prime}, k_z^{\prime} \right> \right|^2,\\
        &\Phi_{N_2 N_1}\left( \xi \right)  =
        \frac{N_2!}{N_1!}\xi^{N_1-N_2}e^{-\xi}\left[ L_{N_2}^{N_1-N_2}\left( \xi \right) \right]^2
    \end{split}
\end{equation}
where $q_{\bot}=\sqrt{q_x^2+q_y^2}$,
$N_1=\max{\left(N, N'\right)}$,
$N_2=\min{\left(N, N'\right)}$,and
$L_{N}^{\alpha}\left(x\right)$ is the Laguerre's polynomial.
Here we also assume that the bloch states are approximated as plane waves
and this is proper for short range impurity potential and weak periodic potential.
For short range impurity potential, $w\left( \mathbf{q} \right)\simeq w$ is independent of $\mathbf{q}$.
In this case, the self-energy in Eq. \eqref{scba_selfenergy} is independent of $\gamma$,
and finally we arrive at\cite{Vasko2005SDH}
\begin{equation}\label{scbaEq}
    \Sigma_{\varepsilon}^{s} = \frac{w}{2\pi l_B^2}\sum_{N, n_b} \int \frac{d k_z}{2\pi}
G^s_{\varepsilon}\left( k_z, N, n_b \right)
\end{equation}

The density of states (DOS) with LB can be directly evaluated
from  Eq.\eqref{green_SCBA} by using
$D_{\text{LB}}\left( \varepsilon \right) =
-g_s\frac{1}{\pi}\sum_{\gamma} \text{Im} G^+_{\varepsilon} \left( \gamma \right)$.
By using  Eqs. \eqref{green_SCBA}, \eqref{scbaEq}, and \eqref{sxy_lb} can be
transformed into\cite{Vasko2005SDH}
\begin{equation}\label{sxxFinalEq}
    \begin{split}
        \sigma_{xx} &= g_s\frac{e^2\omega_c}{8\pi^3}
        \int d \varepsilon \left( -\frac{\partial f\left( \varepsilon \right)}{\partial\varepsilon} \right)
        \frac{\left( 2\Sigma^{\prime\prime} \right)^2}
        {\left( 2\Sigma^{\prime\prime} \right)^2+\left( \hbar \omega_c \right)^2}\\
        &\times\int d k_z\sum_{n_b, N}\frac{\varepsilon-E_{n_b, k_z}-\Sigma^{\prime}}
        {\left( \varepsilon-E_\gamma-\Sigma^{\prime} \right)^2+\left( \Sigma^{\prime\prime} \right)^2},
    \end{split}
\end{equation}
and Eq. \eqref{sxy_lb}
\begin{equation}\label{sxyFinalEq}
    \begin{split}
        \sigma_{xy} &= \frac{e n}{B}-g_s\frac{e^2}{8\pi^3\hbar}
        \int d \varepsilon \left( -\frac{\partial f\left( \varepsilon \right)}{\partial\varepsilon} \right)
        \frac{\left( 2\Sigma^{\prime\prime} \right)^3}
        {\left( 2\Sigma^{\prime\prime} \right)^2+\left( \hbar \omega_c \right)^2}\\
        &\times\int d k_z\sum_{n_b, N}\frac{\varepsilon-E_{n_b, k_z}-\Sigma^{\prime}}
        {\left( \varepsilon-E_\gamma-\Sigma^{\prime} \right)^2+\left( \Sigma^{\prime\prime} \right)^2},
    \end{split}
\end{equation}
where $\Sigma^{\prime\prime}=\text{Im}\Sigma^-_\varepsilon$,
$\Sigma^{\prime}=\text{Re}\Sigma^-_\varepsilon$, and
\begin{equation}\label{nEq}
    n = \int d \varepsilon D_{LB}\left( \varepsilon \right) f\left( \varepsilon \right)
=  \frac{g_s}{\pi w}\int d \varepsilon   \Sigma^{\prime\prime}f\left( \varepsilon \right).
\end{equation}
Eqs. \eqref{sxxFinalEq} and \eqref{sxyFinalEq} are the main results that we will use
later.

\section{Calculations And Results For Kronig-Penney model }
%\subsection{Kronig-Penney model}
As mentioned above, we have formally describe the system.
Through the eigensolution of $h^{\parallel}$,
energy gaps will emerge in the system with periodic potential
due to the Bragg's diffraction\cite{kittel1976introduction}.
\begin{figure}[ptb]
    \centering
    \includegraphics[width=1\columnwidth]{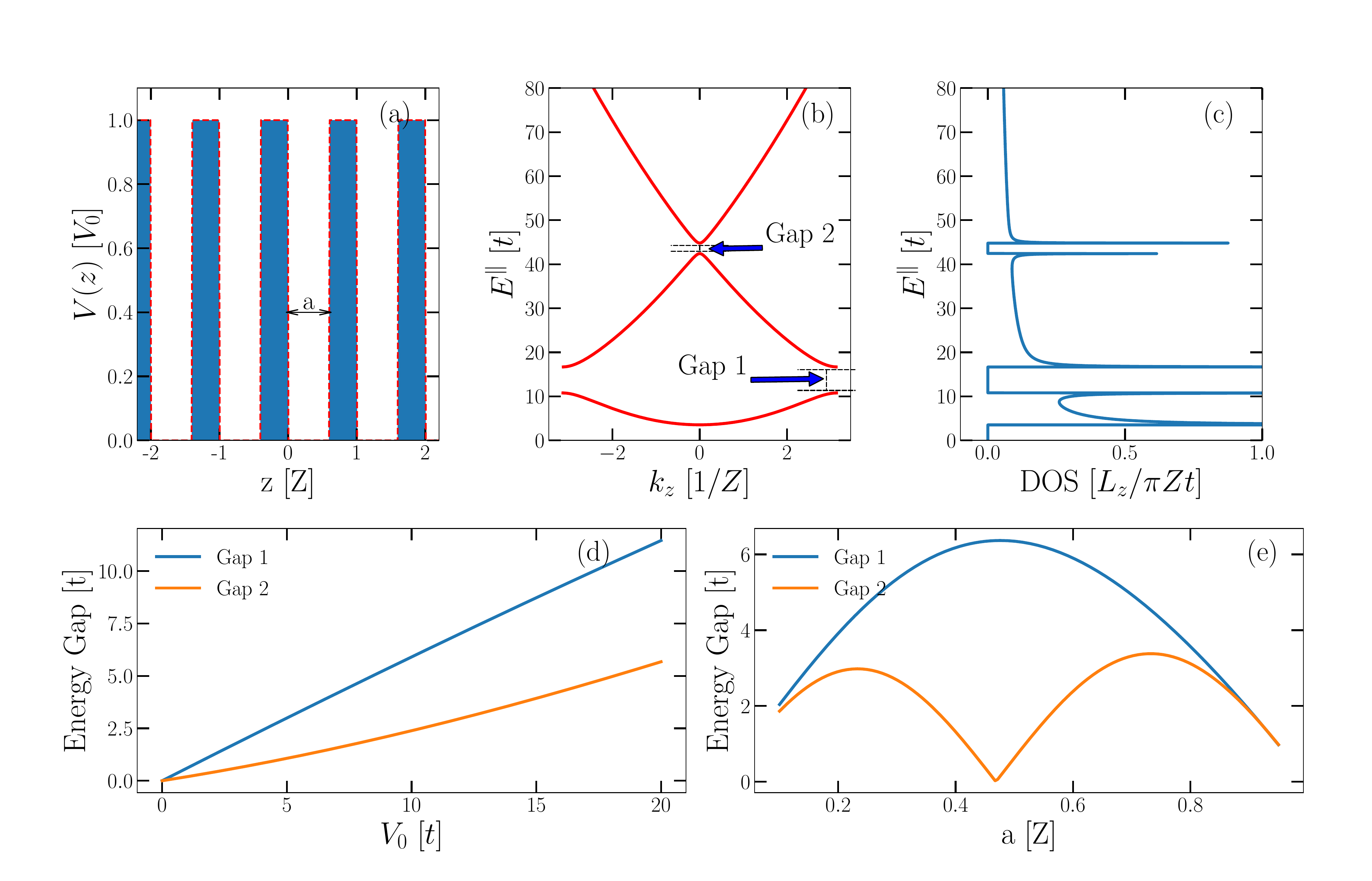}
    \caption{These pictures mainly show the properties of Kronig-Penney model in a $1D$ periodic potential.
    (a)The periodic square-well potential defined in Eq.(\ref{kppot}).
    (b) and (c) The Bloch energy spectrum and the corresponding DOS.
    (d) and (e) The energy gaps as a function of parameters $V_0$ and $a$, respectively.
    Here, we set $a/Z=0.6$, $V_0/t=10$, and $t=\frac{\hbar^2}{2 M Z^2}$.}
    \label{fig:KPMODEL}
\end{figure}
In order to show a specific result,
we start with the Kronig-Penney model with a periodic square-well potential
\begin{equation}\label{kppot}
V\left(z\right)=
    \begin{cases}
    0& ~~ nZ< x <n Z + a\\
    V_0& ~~ nZ+a< z <n Z + a+b
    \end{cases}
    .
\end{equation}
Here, $Z=a+b$ is the barrier period and $V_0$ is the barrier potential as shown in Fig.\ref{fig:KPMODEL}(a).
Using Bloch Theorem $f_{k_z}(z+Z)=e^{i k_z Z}f_{k_z}(z)$,
for eigenvalue $E^{\parallel}$,
we can draw up the eigenfunction
\begin{equation}
f_{k_z}(z) =
\left\{
\begin{split}
& Ae^{iK z}+Be^{-iK z} ~~ 0<z<a\\
& Ce^{Q z}+D e^{-Q z} ~~ -b<z<0 \\
&\left[Ce^{Q (z-Z)}+D e^{-Q (z-Z)}\right]e^{i k_z Z} ~~ a<z<Z
\end{split},
\right.
\end{equation}
where, $E^{\parallel}=\frac{\hbar^2 K^2}{2M}$ and $V_0-E^{\parallel}=\frac{\hbar^2 Q^2}{2M}$.
To get the constant coefficients $A, B, C, D$,
we utilize the boundary conditions that
the wave function $f_{k_z}$ and $df_{k_z}/dz$ are continuous at $z=0 $ and $z=a$, namely,
\begin{equation}\label{matchCondition}
\left\{
\begin{split}
  &A+B =C+D   \\
  &i K \left(A-B\right) = Q\left(C-D\right) \\
  &A e^{i K a}+Be^{-i K a} = \left(Ce^{-Q b} +D e^{Q b}\right)e^{i k_z Z} \\
  &i K \left(A e^{i K a}-Be^{-i K a}\right) =  Q \left(Ce^{-Q b} -D e^{Q b}\right)e^{i k_z Z}
\end{split}.
\right.
\end{equation}
Eq. \eqref{matchCondition} has nontrivial solutions only if
the determinant of the coefficients matrix vanishes, yielding
\begin{equation}\label{bands}
    \frac{Q^2-K^2}{2 Q K}\sinh{Q b} \sin{K a} + \cosh{Q b} \cos{K a} = \cos{k_z Z}.
\end{equation}
% [The energy dispersion relations between $E^{\parallel}_{n_b, k_z}$
% and $k_z$ can be got from solving Eq.(\ref{bands}) numerically as shown in Fig.\ref{fig:KPMODEL}(b),
% where energies disperse in first Brillouin zone.]
Numerically solving Eq. \eqref{bands},
we learn the energy dispersion relations between $E^{\parallel}_{n_b, k_z}$ and $k_z$
in the first Brillouin zone as depicted in Fig.\ref{fig:KPMODEL}(b).
More clearly, the first two energy gaps have been displayed
in Fig.\ref{fig:KPMODEL}(d) and Fig.\ref{fig:KPMODEL}(e).

% [To get the DOS of the 1D Bloch bands,
% we denote left hand side of Eq.(\ref{bands})  as a function
% $f_{LHS}\left(E^{\parallel}\right)$ and the right hand side as $f_{RHS}\left(k_z\right)$.
% By taking the derivation of both sides of  Eq.(\ref{bands}),
% the DOS can be calculated numerically by]
As for the density of states(DOS) of the Bloch bands,
denoting the left and right sides of Eq. \eqref{bands}
as $f_{LHS}\left(E^{\parallel}\right)$ and $f_{RHS}\left(k_z\right)$, respectively,
and taking the derivative of both sides of Eq. \eqref{bands},
we can calculate the DOS numerically by
\begin{equation}\label{BlochDOS}
    D^{\parallel}\left(E^{\parallel}\right) \equiv \frac{L_z}{2\pi} \frac{d k_z}{d E^{\parallel}}
    = \frac{L_z}{\pi Z}
    \frac{\left| f'_{LHS}\left(E^{\parallel}\right)\right| }
    {\sqrt{1- f^2_{LHS}\left(E^{\parallel}\right)}},
\end{equation}
where $f'_{LHS}\left(E^{\parallel}\right) = d f_{LHS} / d E^{\parallel}$.
And it has been plotted in Fig.\ref{fig:KPMODEL}(c).

\begin{figure}[ptb]
    \centering
    \includegraphics[width=1\columnwidth]{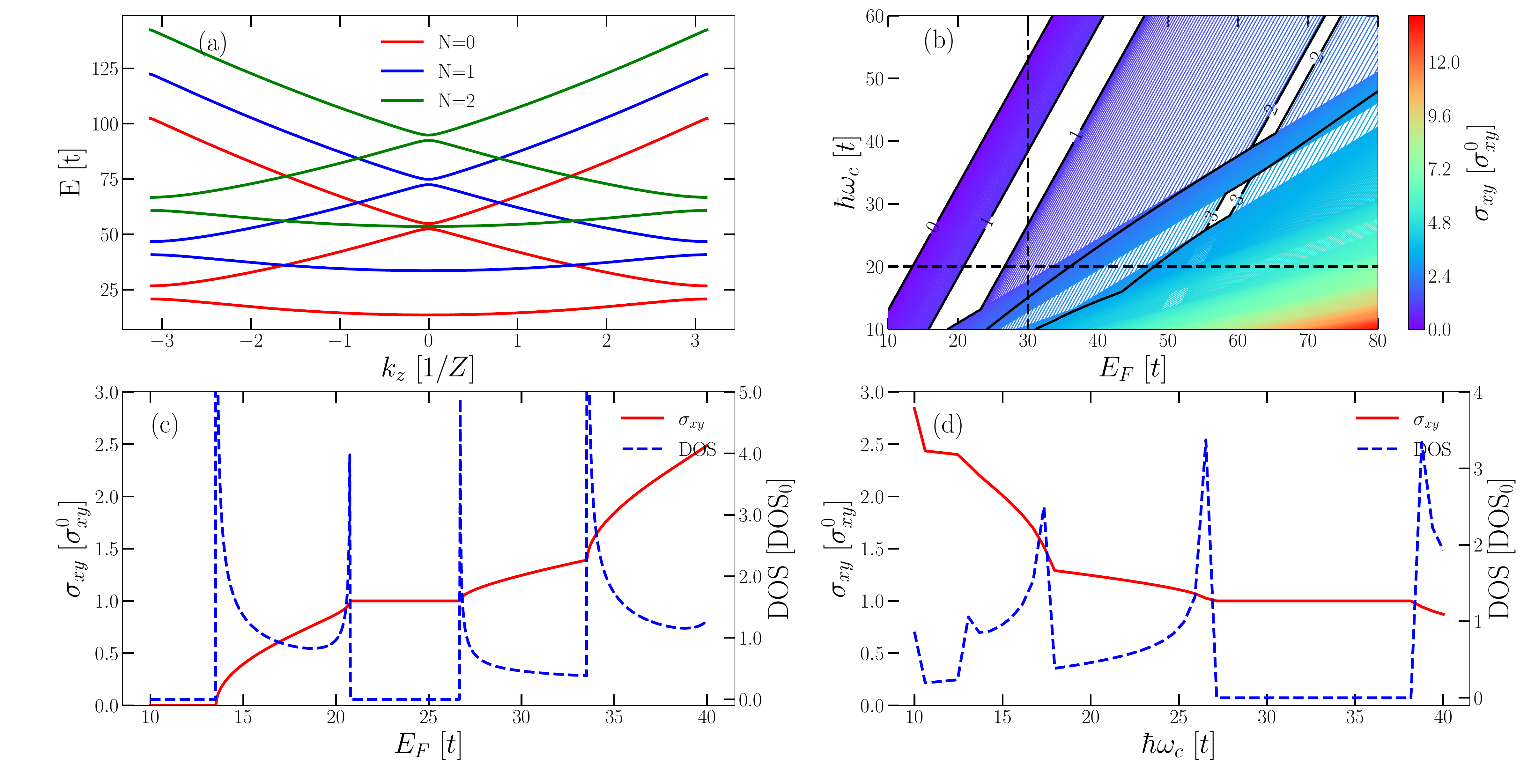}
    \caption{These figures show some results of Landau Level, Hall conductivity
    and corresponding DOS.
    (a)The energy spectrum includes three Landau Levels are plotted
    where we take $\hbar \omega_c/t=20$.
    (b)The contour lines of Hall conductivity in the unit of $\sigma^0_{xy}$
    with varying both $\hbar\omega_c$ and the Fermi level $E_F$.
    It is the blank white space enclosed by black solid contour lines that
    exactly where the Hall plateaus lie
    and the numbers $0,~1,~2,~3$ in black solid contour lines
    denote the quantized values of Hall conductivity.
    The Hall conductivity and the corresponding DOS in the unit of $DOS_0$
    as a function of (c)$E_F$ for $\hbar\omega_c/t=20$
    and (d)$\hbar\omega_c$ for $E_F/t=30$.
    The black dash lines in (b) is corresponding to the cross sections in (c) and (d).
    Here, we set $\sigma^0_{xy}=g_s \frac{e^2}{2\pi h}\frac{2\pi}{Z}$,
    $DOS_0=\frac{\Omega}{\pi Z^3 t}$
    and the other parameters are the same as Fig.1.}
    \label{fig:LLAndHallConds}
\end{figure}
Next, we will study the system described by hamiltonian $H_{0}$.
From the simultaneous equations, Eqs. \eqref{eigenergy} and \eqref{bands},
we obtain the energy bands as plotted in Fig. \ref{fig:LLAndHallConds}(a).
The Hall conductivity from Eq. \eqref{HallCond0} has been calculated
for different Fermi levels and magnetic fields
as painted in Fig. \ref{fig:LLAndHallConds}(b).
Making Eq. \eqref{BlochDOS} go further,
we can get the DOS
\begin{equation}\label{LLDOS}
    D\left(E\right) = \frac{L_xL_y}{2\pi l_B^2}\sum_{N} D^{\parallel}
    \left[E-\left(N+\frac{1}{2}\right)\hbar\omega_c\right],
\end{equation}
The Hall conductivity and the corresponding DOS
have been collected in Fig. \ref{fig:LLAndHallConds}(c) and (d).

\begin{figure}[ptb]
    \centering
    \includegraphics[width=1\columnwidth]{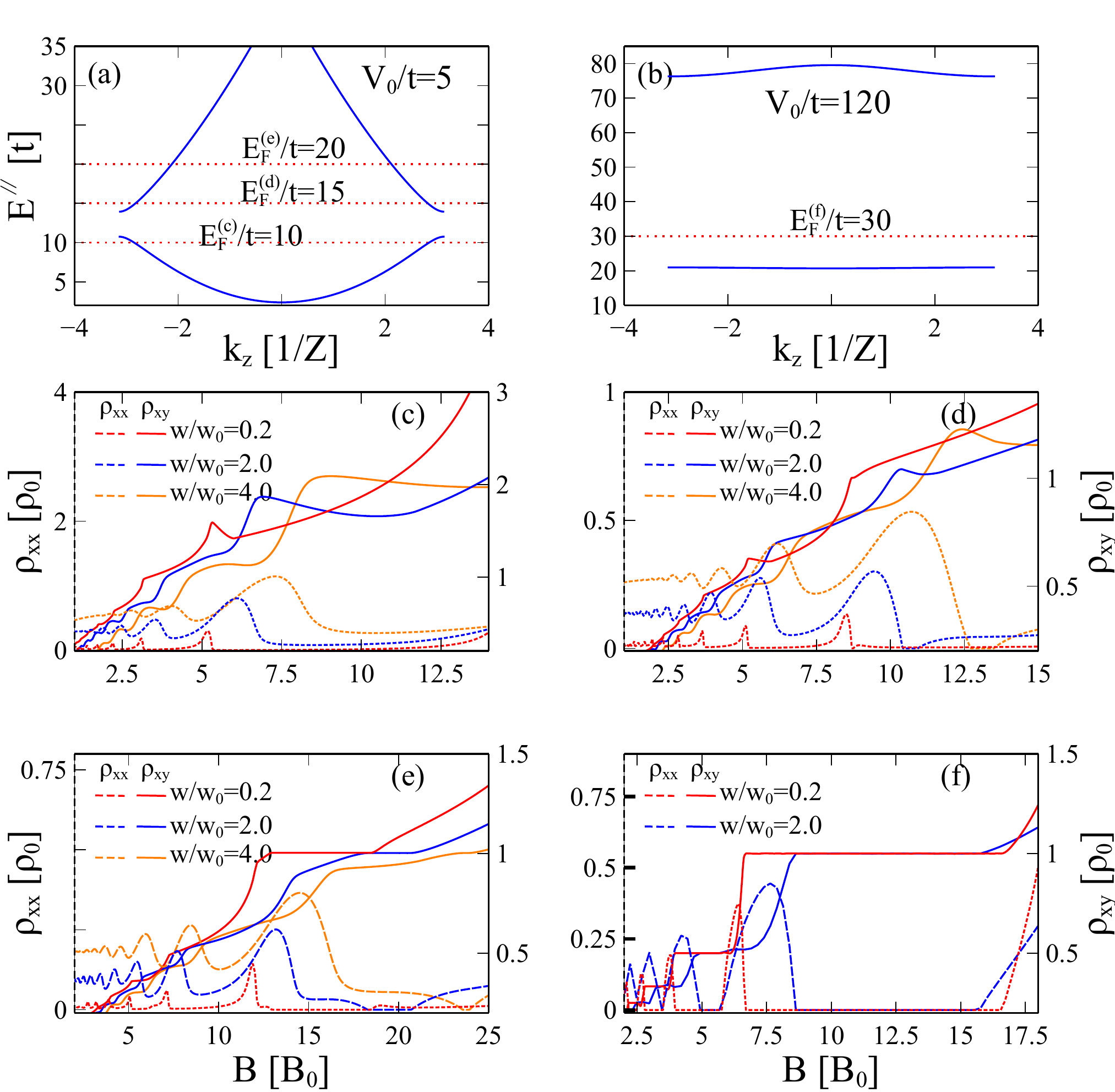}
    \caption{   
    These figures show the influence of the white noise impurities
    on the the Hall resistivities and the TMRs.
        (a), and (b) show show the energy spectrum along $z$ direction
         by given different periodic potential $V_0$.
         In (a), $V_0/t=5$, $a/Z=0.5$, and $E_F^{\text{(c)-(e)}}$ correspond to the Fermi energies
         used in figures (c)-(e) respectively. 
         In (b), $V_0/t=120$, $a/Z=0.5$, and $E_F^{\text{(f)}}$ correspond to the Fermi energy
         used in figure (f). 
        And (c)-(f) show the magneto-resistivities 
         $\rho_{xx}$ and  $\rho_{xy}$ ,
        varying with magnetic field $B$
        at different disorder strengths $w/w_0$ with $w_0= t^2Z^3$.
        The left $y$ axis represents $\rho_{xx}$ and the right $y$ axis represents $\rho_{xy}$.
        Here, we denote $\rho_0=\dfrac{h}{ e^2}\frac{Z}{g_s}$, and 
        $B_0 = \frac{M t}{\hbar e}$.
    }\label{fig:Impurities}
\end{figure}

% [Finally, the effect of impurities are considered which will give rise to  level broadening
% of energy bands, and also result in TMCs.]
By using Eqs. \eqref{green_SCBA}, \eqref{scbaEq}, \eqref{sxxFinalEq}, and \eqref{sxyFinalEq},
TMC and Hall conductivities can be evaluated due to LB.
Furthermore, magneto-resistivities are calculated with relations
\begin{equation}\label{rhoxxEq}
    \rho_{xx} = \frac{\sigma_{xx}}{\sigma_{xx}^2+\sigma_{xy}^2},
\end{equation}
and
\begin{equation}\label{rhoxyEq}
    \rho_{xy} = \frac{\sigma_{xy}}{\sigma_{xx}^2+\sigma_{xy}^2}.
\end{equation}
The results of magneto-resistivities $\rho_{xx}$ and $\rho_{xy}$ varying with magnetic field are shown in Fig. \ref{fig:Impurities},
 Fig. \ref{fig:Impurities}.
For large periodic potential or high Fermi level, 3DQHE retains robust against impurities.
However, when the periodic potential is small and Fermi energies are low, 
$\rho_{xx}$ and $\rho_{xy}$ become subtle.

\section{Discussion And Conclusion}

Now we will continue to make discussions on the above results.
Firstly, as is revealed in Fig. \ref{fig:KPMODEL}(b),
the energy gap can be opened when a periodic potential is applied in $z$ direction.
And with the "strength" of the square-well potential increases,
the gaps increase linearly as Fig. \ref{fig:KPMODEL}(d) given depicts.
Interesting, what Fig. \ref{fig:KPMODEL}(e) presents is that
the gaps may reach an extreme value
when the width of the square-well potential is about half of its period $Z$.

In  Fig. \ref{fig:LLAndHallConds}(a),
we can see that when the parameters, $V_0$ , $a$, and $B$, are chosen properly,
the second Landau band may above the energy appear in Bloch bands in the first Landau band which means robust Hall plateaus may appear.
In Fig. \ref{fig:LLAndHallConds}(b),
we take time to search where the plateaus may appear
by modulating the Fermi energy and magnetic field.
From this figure, we find that
there will be Hall plateaus in the blank region enclosed by the black solid contour lines.
And it's possible that two plateaus show up when the magnetic field and Fermi energy
are both large enough.
Fig. \ref{fig:LLAndHallConds}(c) and Fig. \ref{fig:LLAndHallConds}(d) visualize the corresponding very well.
The DOS vanishes when a Hall plateau appears.
Frome the oscillating DOS,
we can infer that there exists similar oscillations of TMCs.

Besides manifesting the existence of  the 3DQHE, 
the effect of impurities have to be taking into consideration.
As impurities are unavoidable experimentally experimentally,
they have non-negligible effects on transverse magneto-transport which 
is scattering-associated process in fact.
In Figs. \ref{fig:Impurities} (a) and (b), we show the information of the spectrum and Fermi energies,
which are corresponding to  Figs. \ref{fig:Impurities} (c)-(f) respectively.
Figs. \ref{fig:Impurities} (c)-(f) show the TMRs $\rho_{xx}$ and Hall resistivities $\rho_{xy}$,
varies with magnetic field $B$ by taking different impurities strength.
All the results in (c)-(f) are calculated to quantum limit regimes. 
Firstly, from the results we can see that the minimum
TMRs $\rho_{xx}$ always show up in the quantum limit, 
and this results tell us the finite minimum TMRs $\rho_{xx}$ can't be
used as a unique signature of 3DQHE as been observed in 
recent experiments\cite{3dqheExp1, 3dqheExp2,3DQQHEZrTe5, Galeski2020}.
Secondly, (c)-(f) show that $\rho_{xx}$ will increase as the impurities increase, 
and this indicates that the minimum TMRs $\rho_{xx}$ will become finite as the impurities increase
when the periodic potential is not so strong.
In Figs. \ref{fig:Impurities} (c)-(e) energy spectrum remains the same with different typical Fermi energies. 
In Fig. \ref{fig:Impurities} (c), we take the Fermi energy is slightly below the energy gap in the first Bloch band.
In this case, the TMRs and Hall resistivities 
show the similar signatures comparing to the normal 3D electron gas in magnetic field. 
In this regime, the so-called QQHE appears\cite{3DQQHEZrTe5}, 
and as the impurities become stronger, the quasi-quantized plateaus become more like quantized in the 
quantum limit. 
But also the TMRs will become larger.
Fig. \ref{fig:Impurities} (d) shows even more interesting results where we take the Fermi energy slightly 
above the energy gap in the second Bloch band.
The quantized Hall plateaus do not show up which is not so obvious physically.
In this case only QQHE shows, and it's hard to distinguish between
 the cases of periodic system and normal 3D electron gas as in Figs. \ref{fig:Impurities} (c).
This result makes the case even more complex 
that the answer to whether or not 3DQHE can be properly observed experimentally in the 
periodic system becomes unclear.
When the Fermi energy lifts up higher, the 3DQHE will emerge, and is robust to weak impurities.
But even in this regime, things become subtle when the impurities become strong enough as depicted
in Fig. \ref{fig:Impurities} (e) when $w/w_0=4.0$. 
The quantized plateau will evolve into a slope and vanishing TMRs become finite.
This can be understood as an effect of level broadening.
In fact the level broadening $\bar{\varepsilon}$ due to the impurities have a close relation with  $w$,
and it can be  expressed as\cite{Vasko2005} $\bar{\varepsilon}=\hbar/\tau_{sc}=\pi w D_{\text{LB}}$.
After some calculation, we can estimate that $\bar{\varepsilon} /t\sim  5 $ at $w/w_{0} \sim 4$,
and the magnitude of the gap $\delta E /t \sim 5$.
In Fig. \ref{fig:Impurities} (d), we call these regimes as "low" Fermi energy regimes,
and the QQHE appear just because the initial Fermi energy low.
In Fig. \ref{fig:Impurities} (e), the 3DQHE disappear is due to the "strong" impurities comparing 
to the energy gap.

In Fig. \ref{fig:Impurities} (f), we take large periodic potential $V_0$ 
and the results are similar to that of the bulk Hall effect 
in  artificial superlattice system in the article \cite{3DQHEPeriodic1986}.
As a matter of fact, the parameters taking here are mainly estimated from those in the article \cite{3DQHEPeriodic1986},
and we find the artificial superlattice is more likely to construct larger layer periodic barrier potential.
From our results,we see that the larger periodic potential will result in larger energy gap and more robust 3DQHE.
This makes the artificial superlattice an ideal platform for engineering and studying 3DQHE comaring with
the natural materials in which the spontaneous periodic potential is probably weak.

From the above discussions and results,
we conclude that in the periodic system either CDW or SDW-type or superlattice-type periodic potential,
the 3DQHE can appear by tuning the system parameters properly.
A more detail study show that the distinguish between QQHE and 3DQHE
are not so clear just from the signatures of TMRs and Hall resistivities.
The finite minimum values of $\rho_{xx}$ together with quasi-quantized Hall plateaus
can emerge from four different cases.
The first one is the normal 3D electron gas without impurities 
in the vicinity of the quantum limit as showing in the reference \cite{3DQQHEZrTe5}.
The second one is the multi-band explanations as argued in the references \cite{3dqheExp2,Galeski2020}.
We show two new mechanisms, which are the "low" Fermi energy mechanism when Fermi energy is in or slightly
above the Bloch gap and the "strong" impurity effect when the LL broadening is comparable with Bloch band gap.
According to our conclusions, maybe other experimental methods 
should be considered to tell whether the spontaneous periodic exsited in the system,
such as performing the scanning tunneling measurement (STM) in the surface of the $x-z$ or $y-z$ plane
to search for the periodic CDWs, etc..
Also, we argue that the artificial superlattice can be engineered to be an ideal platform for studying 3DQHE properties,
as the large layer barrier periodic potential can be mannually fabricated.

\begin{acknowledgements}
    We really appreciate the useful discussions with R. Ma and W. Luo.
    This work was supported by the State Key Program for Basic Researches of China
    under Grant No. 2017YFA0303203 (D.Y.X.),
    and the National Natural Science Foundation of China under Grant No. 11974168 (L.S.).
\end{acknowledgements}

H. Geng and G. Y. Qi contributed equally to this project.

The code and data used to produce the figures are available online\cite{3dqhecode}.

\bibliographystyle{apsrev4-1}
\bibliography{references}
\end{document}